\documentclass[aip,cha,reprint]{revtex4-1}

\usepackage{color}
\usepackage{graphicx} 
\usepackage{xfrac} 
\usepackage{amssymb, amsmath, amsthm, amsfonts}

\newcommand{\BE}{\begin{equation}}
\newcommand{\EE}{\end{equation}}
\newcommand{\BA}{\begin{eqnarray}}
\newcommand{\EA}{\end{eqnarray}}
\newcommand{\bx}{{\bf x}}
\newcommand{\bv}{{\bf v}}
\newcommand{\mP}{\mathbf{P}}
\newcommand{\mA}{\mathbf{A}}
\newcommand{\mU}{\mathbf{U}}

\begin{document}


\title{Clustering coefficient and periodic orbits in flow networks} 



\author{Victor Rodr\'{\i}guez-M\'{e}ndez}
\affiliation{IFISC (CSIC-UIB), Instituto de F\'{\i}sica
Interdisciplinar y Sistemas Complejos, Campus Universitat de
les Illes Balears, E-07122 Palma de Mallorca, Spain.}
\author{Enrico Ser-Giacomi}
\affiliation{\'{E}cole Normale  Sup\'{e}rieure,  PSL Research
University,  CNRS,   Inserm,  Institut  de  Biologie de
l'\'{E}cole Normale Sup\'{e}rieure (IBENS), F-75005 Paris,
France}
\author{Emilio Hern\'{a}ndez-Garc\'{\i}a}
\affiliation{IFISC (CSIC-UIB), Instituto de F\'{\i}sica
Interdisciplinar y Sistemas Complejos, Campus Universitat de
les Illes Balears, E-07122 Palma de Mallorca, Spain.}


\date{November 2, 2016}

\begin{abstract}
We show that the clustering coefficient, a standard measure in
network theory, when applied to flow networks, i.e. graph
representations of fluid flows in which links between nodes
represent fluid transport between spatial regions, identifies
approximate locations of periodic trajectories in the flow
system. This is true for steady flows and for periodic ones in
which the time interval $\tau$ used to construct the network is
the period of the flow or a multiple of it. In other situations
the clustering coefficient still identifies cyclic motion
between regions of the fluid. Besides the fluid context, these
ideas apply equally well to general dynamical systems. By
varying the value of $\tau$ used to construct the network, a
kind of spectroscopy can be performed so that the observation
of high values of mean clustering at a value of $\tau$ reveals
the presence of periodic orbits of period $3\tau$ which impact
phase space significantly. These results are illustrated with
examples of increasing complexity, namely a steady and a
periodically perturbed model two-dimensional fluid flow, the
three-dimensional Lorenz system, and the turbulent surface flow
obtained from a numerical model of circulation in the
Mediterranean sea.
\end{abstract}


\maketitle 


\begin{quotation}
The Lagrangian description of fluid dynamics, which focuses on
the motion of the fluid particles as they are advected by the
flow, provides a useful bridge between the theory of dynamical
systems and the analysis of fluid transport and mixing, so that
techniques and results can be transferred from one field to the
other. Modern network theory has also been brought into contact
with fluid dynamics and dynamical systems through the concept
of flow networks, in which the motion of fluid particles
between different regions is represented by links in a graph.
In this paper we use the flow network framework to show that
the clustering coefficient, a standard measure in network
theory, identifies periodic orbits, fundamental objects in the
theory of dynamical systems and also of importance in the
context of fluid motion.
\end{quotation}


%
%

%

\section{Introduction}
\label{sec:intro}

In the last years flow networks have been defined and analyzed
to understand properties of fluid transport and mixing
\cite{sergiacomi2015flow,sergiacomi2015most,sergiacomi2015dominant,lindner2017spatio},
with particular emphasis on geophysical transport and its
biological implications
\cite{treml2008,jacobi2012identification,rossi2014hydrodynamic,dubois2016linking}.
The basis of the flow-network paradigm rests on addressing the
Lagrangian fluid dynamics of the system with the so-called
set-oriented methods
\cite{Froyland2003,Froyland2005,Froyland2007,Dellnitz2009,
Froyland2010,Santitissadeekorn2010,Froyland2012Rossi,Tallapragada2013,Bolltbook2013}
in which a discrete approximation to the motion of fluid is
characterized by a transfer or Perron-Frobenius operator
indicating which proportion of fluid is moved from one region
to another by the flow. Under the name of `mapping method'
these ideas have provided insight on mixing and its
optimization in the context of industrial flows
\cite{galaktionov2002mapping,singh2008optimizing,singh2008mapping}.
Besides the context of fluid flows, the framework is also
relevant to characterize the behavior of general dynamical
systems
\cite{froyland2003detecting,dellnitz2006graph,santitissadeekorn2007identifying,Bolltbook2013}.
The contact with network theory \cite{Newman2010} is made when
the matrix representing the transfer operator is interpreted as
the adjacency matrix of a network
\cite{dellnitz2006graph,santitissadeekorn2007identifying,Bolltbook2013,
sergiacomi2015flow,sergiacomi2015most,sergiacomi2015dominant,lindner2017spatio},
so that the weight of a link between two nodes is given by the
amount of flow between the corresponding spatial locations. An
alternative viewpoint in characterizing fluid flows with
network techniques assigns links between spatial regions
according to statistical correlations between their dynamical
variables, leading to correlation-based flow networks
\cite{molkenthin2014a,tupikina2016correlation,molkenthin2017edge}
and making a connection with the broad field of climate
networks
\cite{tsonis2004,yamasaki2008climate,donges2009a,barreiro2011}.

Flow networks have been analyzed with standard measures of
network topology, which have complemented the direct study with
specific fluid dynamics methods. For example, degree and
related properties of the different nodes have been related to
local mixing and dispersion, with explicit correspondence with
Lyapunov exponents and entropic measures
\cite{sergiacomi2015flow}. Betweenness centrality highlights
preferred transit nodes connecting distant regions
\cite{sergiacomi2015most}. Closeness and eigenvector centrality
distinguish regions dominated by laminar or by strong mixing,
and identify structures related to invariant manifolds
\cite{lindner2017spatio}. Spectral and other community
detection methods have been used to extract coherent regions
and basins of attractions in the fluid flow
\cite{froyland2007detection,froyland2010transport,jacobi2012identification,froyland2014how,
rossi2014hydrodynamic,sergiacomi2015flow}.

There is however a standard network measure, the
\emph{clustering coefficient}, which has not yet been
interpreted in the context of fluid flows. The aim of the
present paper is to provide such interpretation. We show that,
under some conditions, clustering characterizes in a useful way
periodic trajectories in fluids and in dynamical systems,
serving as a spectroscopic tool that detects the presence of
periodic or close-to-periodic fluid paths of given period, and
reveals their approximate location in space. Periodic orbits
are objects of fundamental importance in the theory of
dynamical systems\cite{ChaosBook} and in the context of fluid
flow they identify recurrent fluid-element motions. Even when
the conditions required to locate periodic trajectories are not
perfectly fulfilled, the clustering coefficient still reveals
interesting flow structures.

To achieve our aim we will consider examples representative of
flows with an increasing level of complexity. After revising
the methodology of network construction (Sect.
\ref{sec:flownet}) and discussing the general relationship
between clustering and closed paths in networks (Sect.
\ref{sec:clustering}), we address in Subsect.
\ref{subsec:steady2d} the simple situation of two-dimensional
(2d) steady flows. In this case the associated flow network is
also steady and the clustering measures have a clear meaning.
The interpretation of these quantities are discussed for 2d
periodic flows in Subsect. \ref{subsec:periodic2d} and for a
three-dimensional (3d) time-independent dynamical system in
Subsect. \ref{subsec:steady3d}. A situation of fully aperiodic
flow is briefly addressed in Subsect. \ref{subsec:aperiodic}.
The paper closes with a sections mentioning further extensions
(Sect. \ref{sec:extensions}) and the Conclusions section (Sect.
\ref{sec:conclusions}).

\section{Flow network construction from fluid motion}
\label{sec:flownet}

Fluid flow is a process occurring in continuous space. A
network representation of it requires some kind of
discretization or {\sl coarse-graining}. We briefly remind here
the basic steps needed to build discrete flow networks from the
Lagrangian fluid
dynamics\cite{sergiacomi2015flow,lindner2017spatio}.

First, the spatial domain occupied by the fluid is discretized
in a large number $N$ of boxes, $\{B_i, i=1,...,N\}$, so that
network node $j$ will represent the fluid box $B_j$. Although
this is not strictly necessary, we take here for simplicity all
boxes to have the same area or volume. Also, unless otherwise
stated, we consider an incompressible flow of constant density.
Then, all boxes contain the same amount of fluid.

To complete the definition of the transport network, we
establish a directional \emph{link} between two nodes when an
exchange of fluid occurred between the corresponding fluid
boxes during a given time interval. To do this, we use the
fluid {\sl flow map} $\Phi_{t_0}^\tau$:
\BE
\bx(t_0+\tau)=\Phi_{t_0}^\tau (\bx_0)
\label{map}
\EE
which gives the position at time $t_0+\tau$ of the fluid
particle started at $\bx_0$ at time $t_0$, and that is obtained
simply by integrating the equations of motion of the fluid
elements in the velocity field $\dot \bx(t)=\bv(\bx(t),t)$. By
considering this Lagrangian motion of all fluid particles
inside a region $A$ we define the action of the fluid map on
whole regions: $A(t_0+\tau)=\Phi_{t_0}^\tau (A(t_0))$. By
applying the flow map to our discrete boxes, we obtain the
amount of flow among each pair of nodes, which will be taken as
the {\sl weight} of the corresponding link. More explicitly,
the fraction of the fluid that started at node $i$ at time
$t_0$ and ended up at node $j$ after a time $\tau$ is given by
\cite{Froyland2003,Froyland2005,Froyland2007,Dellnitz2009,Froyland2010,Santitissadeekorn2010}:
\BE
\mP(t_0,\tau)_{ij} = \frac{m\left(B_i \cap
\Phi_{t_0+\tau}^{-\tau}(B_j)\right)}{m(B_i)} \ . \label{PF}
\EE
The transfer or transport matrix $\mP(t_0,\tau)_{ij}$ is a
discrete approximation to the Perron-Frobenius operator of the
flow. $m(A)$ is a measure assigned to the set $A$, which in our
case is its area or volume (and, for constant density, the
amount of fluid it contains). A probabilistic interpretation of
Eq. (\ref{PF}) is that $\mP(t_0,\tau)_{ij}$ is the probability
for a particle to reach the box $B_j$, under the condition that
it started from a uniformly random position within box $B_i$.
We interpret $\mP(t_0,\tau)$ as the adjacency matrix of a
weighted and directed network, so that $\mP(t_0,\tau)_{ij}$ is
the weight of the link from node $i$ to node $j$, for given
$t_0$ and $\tau$. $\mP(t_0,\tau)$ is row-stochastic, i.e. it
has non-negative elements and $\sum_{j=1}^{N}
\mP(t_0,\tau)_{ij}=1$. For incompressible flows it is also
doubly stochastic.

A standard numerical estimation of (\ref{PF}) is obtained by
randomly placing at time $t_0$ a large number of particles
$N_i$ inside each box $B_i$ (with our equal area/volume choice,
all $N_i$ should be equal), and determine with the flow map in
which boxes are the final positions at time $t_0+\tau$. Then:
\BE
\mP(t_0,\tau)_{ij} \approx \frac{\textrm{number of particles
from box $i$ to box $j$}}{N_i} \ . \label{PFparticles}
\EE

Networks constructed from $\mP(t_0,\tau)$ are \emph{weighted}
and \emph{directed} \cite{Newman2010}. In addition, the
dependence on $t_0$ gives in fact a different network for each
starting time $t_0$. We can consider the sequence
$\{\mP(t,\tau), t=t_0,t_1,...\}$ as a \emph{temporal} network
\cite{holme2012temporal}, i.e. a network which is changing in
time. The case in which $\mP(t_0,\tau)$ is independent on the
initial time $t_0$ gives a \emph{static} network. This occurs
when the flow velocity field is steady: $\bv(\bx,t)=\bv(\bx)$.

It is
standard\cite{dellnitz2009seasonal,Froyland2012Rossi,froyland2014how}
to make a Markovian approximation for the dynamics at times
beyond $\tau$: we assume that the transport matrix at long
times is well approximated by a product of matrices at shorter
times, $\mP(t_0,n\tau)=\mP(t_0,\tau)\mP(t_0+\tau,\tau) ...
\mP(t_0+(n-1)\tau,\tau)$. This amounts to assuming that at the
beginning of each time interval the fluid particles are
reinitialized with uniform density in each box, losing memory
of their earlier positions. The assumption is equivalent to
introducing some artificial diffusion in the dynamics
\cite{froyland2013analytic}. In the limit of very small boxes
and very short time steps, the approximation becomes
correct\cite{froyland2013analytic}, as this computational
diffusion vanishes and we recover the exact Lagrangian dynamics
given by (\ref{map}). For finite boxes the added diffusion can
be thought as modeling unresolved scales of motion.

A path in a network is an ordered set of nodes $\{i,j,k,...\}$
such that there is a non-vanishing link between each
consecutive pair of nodes. A \emph{time-respecting
path}\cite{holme2012temporal} can be defined for a temporal
network $\{\mP(t,\tau), t=t_0,t_1,...\}$ as a sequence of nodes
$\{i,j,k,...\}$ such that there is a non-vanishing link between
$i$ and $j$ for the first time interval $[t_0,t_1]$, a
non-vanishing link between $j$ and $k$ for the second interval
$[t_1,t_2]$, etc. Time-respecting paths characterize transport
in the flow network
\cite{lentz2013unfolding,sergiacomi2015most,sergiacomi2015dominant}.
Since nodes are finite regions of flow, time-respecting paths
can be though as `tubes' of fluid trajectories giving a
coarse-grained version of them. When no confusion could arise
we would refer to time-respecting paths also as
\emph{trajectories}. Of course, for static networks there is no
distinction between paths and time-respecting paths.

Many properties of networks defined by the transport matrix
$\mP(t_0,\tau)$ have been already analyzed, as summarized in
the Introduction. We now consider the clustering coefficient.

\section{Clustering coefficient and closed paths}
\label{sec:clustering}

\begin{figure}
\centering
\includegraphics[width=.8\columnwidth, clip=true]{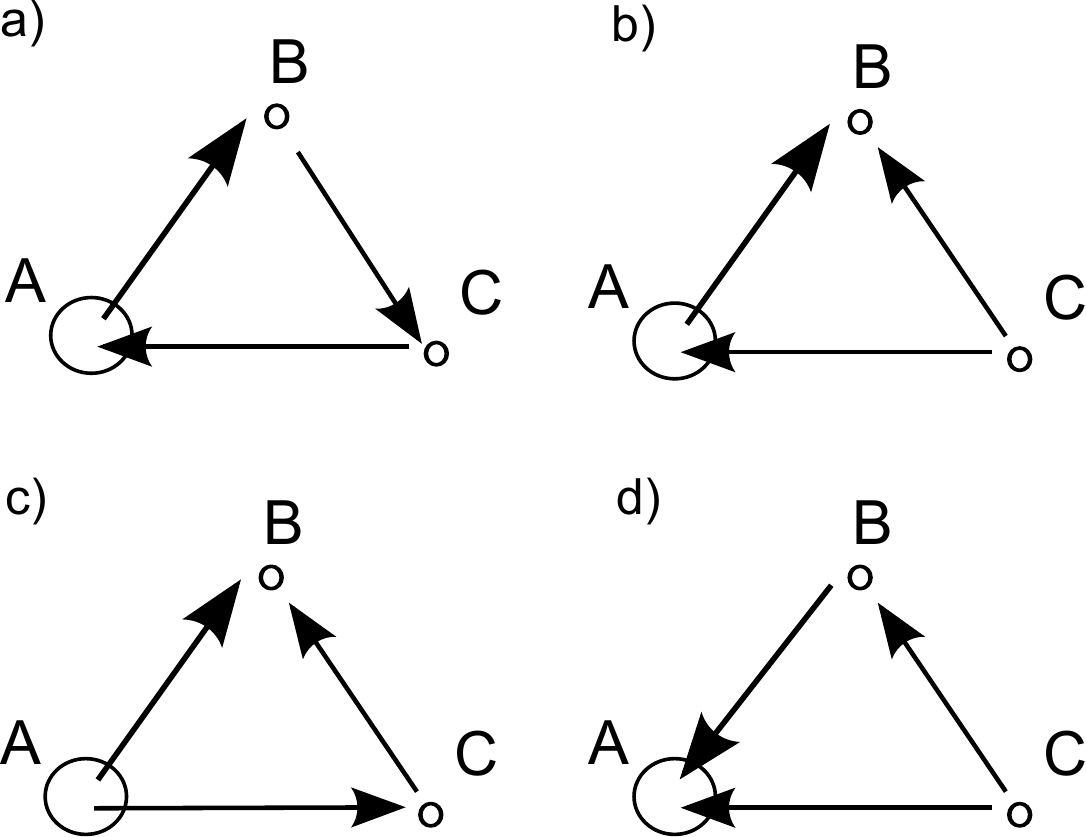}
\caption{Four distinct configurations of the directions of the edges in a triangle involving
nodes $A$, $B$ and $C$. Case a) is a cyclic triangle. Node $A$ in configuration
c) acts as a source, whereas it is a sink in configuration d). $A$ is just an intermediate
node in configuration b). } \label{fig:triangles}
\end{figure}

The clustering coefficient of a node measures the amount of
closed triangles in the network of which that node is a vertex.
\cite{newman2003structure,newman2009networks}. Depending on the
type of network (weighted, directed, ...) and of the kind of
triangles one is interested in, different clustering
coefficients can be
defined\cite{saramaki2007generalizations,fagiolo2007clustering}.
Triangles in a graph give closed paths and then it is natural
to look for relationships between these triangles and closed
trajectories of the fluid elements, the signature of periodic
orbits of the associated advection dynamical system. An
essential characteristic of the fluid trajectories is its
directionality. Thus, considering the \emph{directed} character
of the flow network would be an important ingredient to take
into account when defining a clustering coefficient of
relevance in the study of fluid flows. To focus more clearly on
this relevant aspect we will neglect in the following the
weighted character of flow networks, by considering the
unweighted network defined by the adjacency matrices
$\mA(t_0,\tau)$, binary versions of the transport matrices
$\mP(t_0,\tau)$:
\BA
\mA(t_0,\tau)_{ij}=1 \ \ \ &\textrm{if} \ &\mP(t_0,\tau)_{ij}>0
 \  \textrm{and} \  i \neq j\nonumber \\
\mA(t_0,\tau)_{ij}=0 \ \ \ &\textrm{if} \
&\mP(t_0,\tau)_{ij}=0\
 \  \textrm{or} \  i=j \ . \label{binary}
\EA
$\mA(t_0,\tau)$ still keeps the connectivity and directional
information, as well as the time-evolving character through the
time-dependence on $t_0$. Let us consider first the single
network defined by $\mA(t_0,\tau)$ for fixed $t_0$. Not all
triangles present there are related to closed paths. Figure
\ref{fig:triangles} shows four possible configurations of the
directions in the edges of a triangle involving nodes $A$, $B$
and $C$. Only configuration a), which is called a
\textit{cyclic} triangle \cite{fagiolo2007clustering}, gives a
triangle that can be related to a closed path $A \rightarrow B
\rightarrow C \rightarrow A$. Given a node $i$ with in-degree
$k_i^{IN}=\sum_{j\ne i} \mA_{ji}$, out-degree
$k_i^{OUT}=\sum_{j\ne i} \mA_{ij}$ and with $k_i^B$ of these
links being bidirectional ($k_i^B=\sum_{j\ne i} \mA_{ij}
\mA_{ji}$), a \emph{cyclic clustering coefficient} $C^c_i$ is
defined as the ratio of all cyclic triangles involving node $i$
present in the network, divided by all possible cyclic
triangles that could have been constructed with these values of
$k_i^{IN}$, $k_i^{OUT}$, $k_i^B$. It can be
computed\cite{fagiolo2007clustering} from the diagonal elements
of the third power of the adjacency matrix $\mA$ :
\BE
C^c_i = \frac{\left(\mA^3\right)_{ii}}{k_i^{IN}
k_i^{OUT}-k_i^B} \ . \label{cc}
\EE

With the particular normalization in Eq. (\ref{cc}), $C_i^c \in
[0,1]$. But, since we have already neglected weight
information, we will not pay much attention to the particular
value of $C_i^c$. Rather the interesting point is if it is zero
or not. In the last case there is at least one directed
triangle involving $i$ in the network. What is the meaning of
this when the network represents a fluid flow, with binary
adjacency matrix $\mA(t_0,\tau)$? It means that there are at
least two intermediate nodes $m$, $n$ such that during the time
interval $[t_0,t_0+\tau]$, some fluid particles have been
transported from $i$ to $m$, some from $m$ to $n$, and some
from $n$ to $i$. Note that then, the values of $C_m^c$ and
$C_n^c$ are also non-zero.

In the case of a steady flow, the flow network will be
time-independent, and the dependence on the initial time $t_0$
in the transport and adjacency matrices will disappear:
$\mP(t_0,\tau)=\mP(\tau)$, $\mA(t_0,\tau)=\mA(\tau)$. In this
situation (under the Markovian assumption mentioned above),
there will be fluid particles actually following the cycle
$i\rightarrow m \rightarrow n$ during a time interval of
duration $3\tau$. Nodes $i$ with non-vanishing $C_i^c$ will be
part of time-respecting paths, i.e. coarse-grained fluid
trajectories, of period $3\tau$. Since fluid trajectories are
continuous and different points on them can be associated with
different values of $t_0$, the invariance of $\mA$ with respect
to $t_0$ implies that the nodes with non-vanishing clustering
will arrange in (thick) lines providing approximations to
periodic orbits of period $3\tau$.

The situation is very similar if the flow is periodic with the
same period $\tau$ as the duration of the time-interval used to
construct the network (or a submultiple of it). In this case
$\mP(t_0+\tau,\tau)=\mP(t_0,\tau)$ (the same is valid for
$\mA(t_0,\tau)$) and the repeated action at consecutive times
of the transport process is given by powers of a single matrix
$\mP(t_0,\tau)$. At variance with the static case, the
dependence on $t_0$ remains (and it is periodic). Nodes of high
clustering will again give an approximation to the position at
time $t_0$ of periodic orbits of period $3\tau$. In general
these nodes will not form continuous lines, instead the nodes
with high clustering will move following the full trajectory by
changing $t_0$.

If the integration time $\tau$ defining the flow network does
not coincide with the period of the flow (or a multiple of it)
the nodes with a high clustering coefficient are more difficult
to interpret. It is still true that there is motion of fluid
particles between nodes $i\rightarrow n \rightarrow m$ during
the time interval $[t_0+\tau,\tau]$. But after this time the
new transport matrix $\mP(t_0+\tau,\tau)$ will be different
from $\mP(t_0,\tau)$ and then there is no guarantee that the
same motion will repeat three times to produce a periodic
time-respecting path of period $3\tau$. The situation is indeed
similar to that of a fully aperiodic or turbulent flow, which
keeps an arbitrary $t_0$ dependence on $\mP(t_0,\tau)$ and
$\mA(t_0,\tau)$.

\section{Application to specific flows}
\label{sec:specific}

\subsection{Steady two-dimensional flow} \label{subsec:steady2d}

\begin{figure*}
\centering
\includegraphics[width=\textwidth, clip=true]{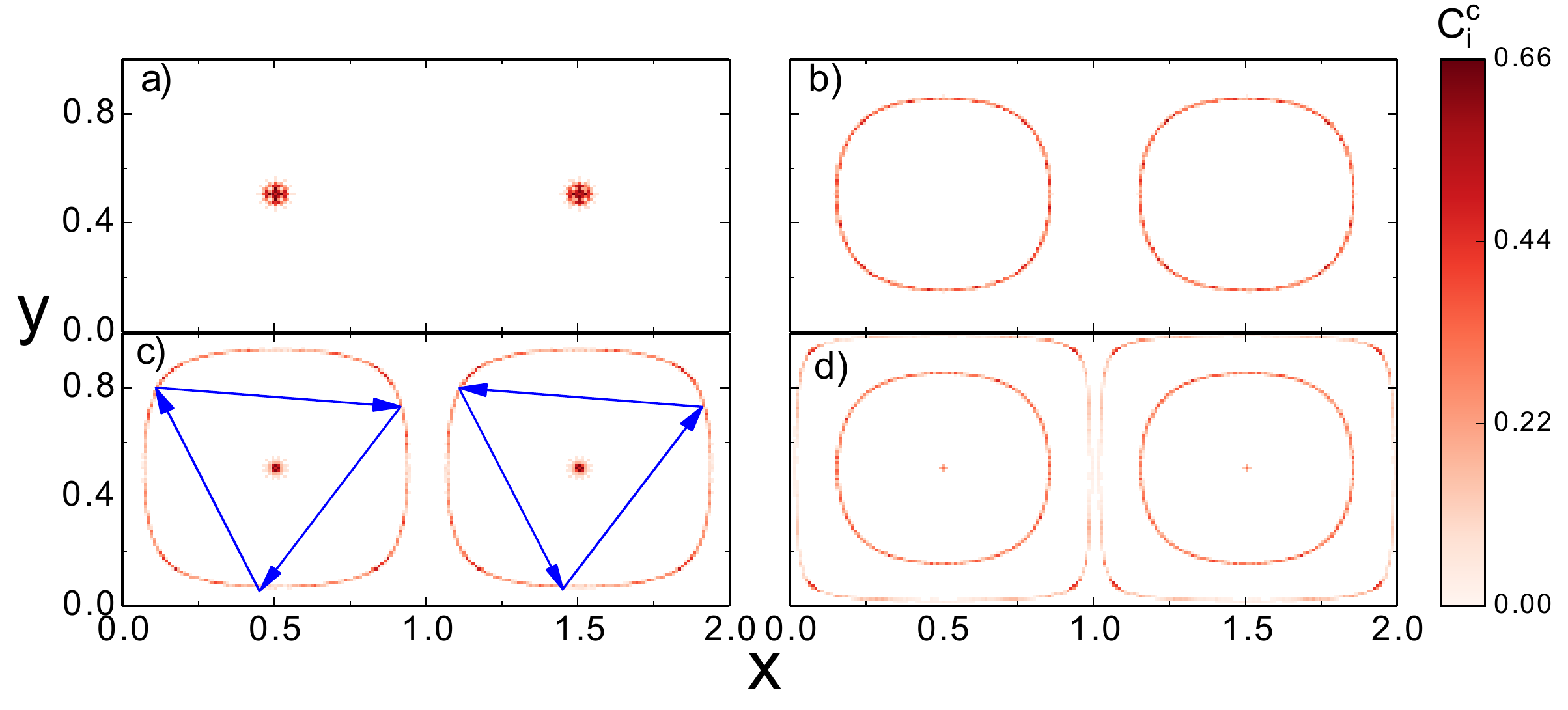}
\caption{Clustering coefficient $C_i^c$ in the nodes of the
flow network constructed from the steady ($\epsilon=0$) double
gyre flow with $A=0.1$. a) $\tau=2$, b) $\tau=3$, c) $\tau=4$,
d) $\tau=5$. Panel c) also shows two examples of cyclic
triangles present in the system. } \label{fig:steady}
\end{figure*}

In this section and in the following we consider an analytic
model flow, the \emph{double-gyre}, which provides a convenient
workbench to relate fluid dynamics quantities to network
characteristics. See for example
\cite{shadden2005definition,farazmand2012computing} for basic
properties of this system and computations of its Lagrangian
coherent structures and Lyapunov fields. Flow networks have
already been constructed from it
\cite{sergiacomi2015dominant,lindner2017spatio}. The
double-gyre is a two-dimensional time-periodic flow defined in
the rectangular region of the plane, $\bx=(x,y) \in
[0,2]\times[0,1]$. It is described by the streamfunction
\BE
\psi(x,y,t)= A \sin(\pi f(x,t))\sin(\pi y) \ ,
\label{dg-stream1}
\EE
with
\BA
f(x,t)&=& a(t)x^2+b(t)x \\
a(t)  &=& \epsilon \sin(\omega t) \ , \\
b(t)  &=& 1-2\epsilon \sin(\omega t) \ . \label{dg-stream2}
\EA
From these expressions, the velocity field is
\BA
\dot x &=& -\frac{\partial\psi}{\partial y}=-\pi A \sin(\pi f(x,t))\cos(\pi y)\\
\dot y &=& \frac{\partial\psi}{\partial x}=\pi A \cos(\pi
f(x,t)) \sin(\pi y) \frac{\partial f(x,t)}{\partial x} \ .
\label{dg-v}
\EA
We take $A=0.1$. We consider in this subsection $\epsilon=0$,
for which the flow is steady, $f(x,t)=x$, and the terms
containing the periodic forcing are absent. This provides a
simple test case to check the behavior of the clustering
coefficient. Ideal fluid particles follow very simple
trajectories: they rotate following closed streamlines,
clockwise in the left half of the rectangle, and
counterclockwise in the right one. The central streamline
$x=1$, a heteroclinic connection between the hyperbolic point
at $(1,1)$ and the one at $(1,0)$, acts as a separatrix between
the two regions. There are hyperbolic fixed points at
$(x,y)=(0,0),(1,0),(2,0),(0,1),(1,1),(1,2)$ and elliptic fixed
points at $(x,y)=(0.5,0.5)$ and $(1.5,0.5)$. The frequency of
the orbits close to the elliptic points can be obtained by
linearization, giving $\omega_0=\pi^2 A$, from which the period
is $T_0=2/(\pi A)\approx 6.366$. The orbits arrange
concentrically around each elliptic point, with period
increasing from $T_0$ when increasing the distance to the
elliptic point and diverging for the nearly-square orbits that
are close to touch the domain walls and the central $x=1$
separatrix.

We discretize the fluid domain into $100\times 50 = 5000$
square boxes, so defining the nodes in our flow network (each
node thus represents a square region of size $0.02\times
0.02$). The transport matrix $\mP(t_0,\tau)=\mP(\tau)$ is
computed by releasing 400 particles from each of the boxes. We
analyze the resulting clustering of each node for different
values of $\tau$.

Figure \ref{fig:steady} clearly shows how the clustering
coefficient computed from a flow network of interval $\tau$
highlights the periodic orbits of period close to $3\tau$:  For
small $\tau$ (not shown) clustering is essentially zero
everywhere. When $\tau=2$ clustering begins to be nonzero for
nodes close to the origin, revealing the presence there of
periodic orbits of period approximately $3\times 2=6$. The
exact linear period close to the origin is $T_0=6.366$,
increasing outwards. This kind of `resonance' at $3\tau=6$ is
broad, revealing the orbits for $\tau$ somehow larger and
smaller than the one corresponding to the exact value
$T_0/3=2.122$. When increasing $\tau$, orbits further from the
origin are highlighted, corresponding to the larger period. For
$\tau$ sufficiently large, orbits with period submultiple of
$3\tau$ are also detected (the origin again for $\tau\approx
4$, a second orbit with period close to $7.5$ for $\tau=5$,
etc.)

\subsection{Periodic two-dimensional flow}
\label{subsec:periodic2d}

\begin{figure}
\centering
\includegraphics[width=\columnwidth, clip=true]{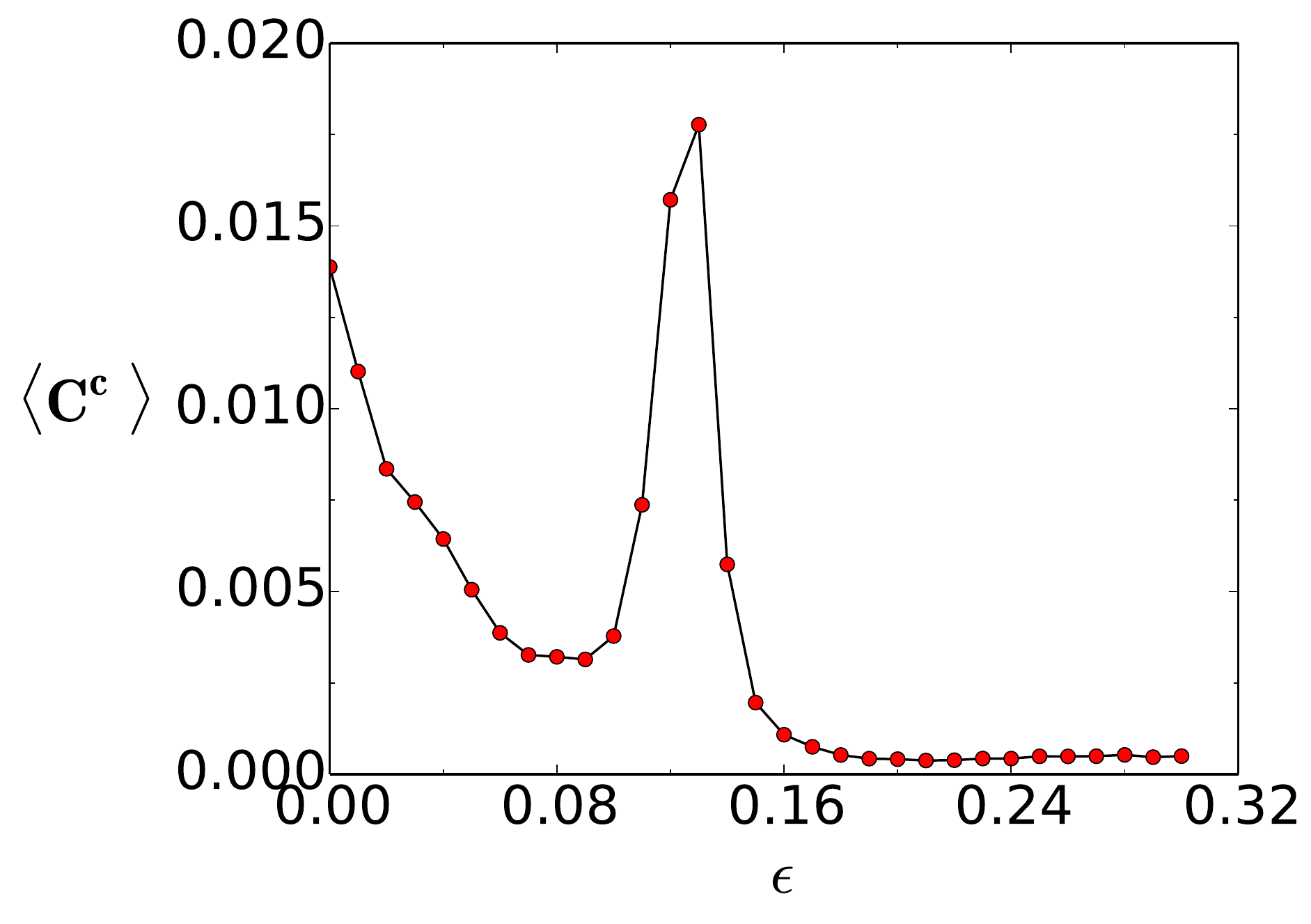}
\caption{Mean clustering coefficient per node $\left< C^c \right>$ as a function of forcing intensity
$\epsilon$ for the flow network with $A=0.1$, $t_0=0$ and $\tau=5$ equal to the flow
period $2\pi/\omega$.}
\label{fig:SpectTau5}
\end{figure}

\begin{figure*}
\centering
\includegraphics[width=\textwidth, clip=true]{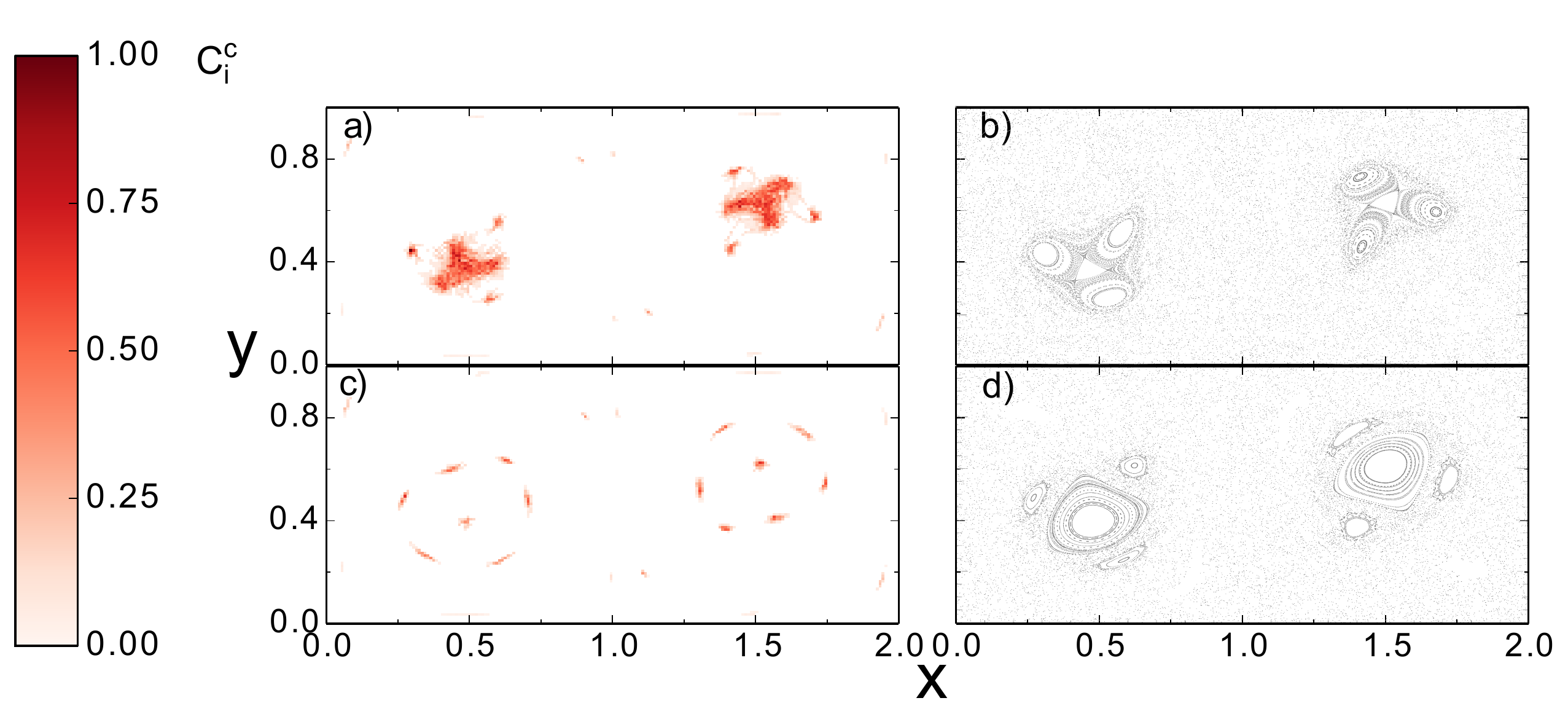}
\caption{Clustering coefficient (panels a) and c)) and
Poincar\'{e} section at $t_0=0$ (panels b) and d)) for the flow
network constructed from the double gyre with $A=0.1$ and
$2\pi/\omega=\tau=5$. Panels a) and b): $\epsilon=0.12$. Panels
c) and d): $\epsilon=0.09$.}
\label{fig:periodic}
\end{figure*}

We consider now the periodically forced double gyre, i.e. Eqs.
(\ref{dg-stream1})-(\ref{dg-v}) with $\epsilon>0$. The
remaining parameters are the same as in the previous subsection
with the addition of the forcing frequency, which we take to be
$\omega=2\pi/5$ so that the flow period is $T=5$, and the
initial time that we always take as $t_0=0$. The procedure of
discretization and particle releasing needed to define the flow
network are also the same as in the previous Subsect.
\ref{subsec:steady2d}. When $\epsilon > 0$, complex behavior
including chaotic trajectories arises. The periodic
perturbation breaks the central separatrix and now some
interchange of fluid is possible between the left and the right
parts of the domain. The geometric structures involved in this
interchange have been studied with a variety of techniques
\cite{shadden2005definition,farazmand2012computing}.

The cleanest interpretation of the clustering coefficient is
obtained when the flow network is computed during one period of
the forcing, i.e. for $\tau=T=5$, so that
$\mP(t_0+\tau,\tau)=\mP(t_0,\tau)$ and a single matrix (and
thus a single network) describes the full dynamics. In this
case clustering is expected to be high in the regions
surrounding the locations of the orbits of period $3\tau=3T$,
i.e. the period-3 orbits. Submultiples of this period will also
lead to non-vanishing clustering, which includes the case of
period-1 orbits. The exact period-1 cycle is indeed excluded,
because self-loops into a node are eliminated by the definition
of the adjacency matrix $\mA(t_0,\tau)$ in Eq. (\ref{binary}),
but still we expect the tube of trajectories surrounding the
periodic one, arising from the coarse-grained nature of the
network description, to contain paths that connect to
neighboring nodes and return some mass to the original one
after three iterations, giving non-vanishing clustering.

In order to monitor the changes in the clustering spatial
distributions as parameters are varied, we compute a global
quantity, $\left<C^c\right>$, the mean clustering per node. We
display it in Fig. \ref{fig:SpectTau5} as a function of
$\epsilon$. We see a marked peak occurring close to
$\epsilon\approx 0.13$. Figure \ref{fig:periodic} shows the
clustering coefficient of the different nodes in the network
(panel a)) for $\epsilon=0.12$, a value close to the maximum of
$\left<C^c\right>$, and compares it with the Poincar\'{e}
section (panel b)) of the same flow at the same time $t_0=0$.
We see that the dominant features in both the Poincar\'{e}
representation and the clustering spatial distribution are the
locations at $t_0=0$ of two period-1 orbits, each one
surrounded by the three elements of an elliptic period-3 orbit
(separated by the elements of the corresponding hyperbolic
orbits of period-3, which have also non-zero $C^c_i$). For
other parameters of the forcing these orbits are also present,
as shown in Fig. \ref{fig:periodic}c and d. Its footprint in
the coarse-grained phase space defined by the network is
however weaker as indicated by the smaller mean
$\left<C^c\right>$ in Fig. \ref{fig:SpectTau5}.

\begin{figure}
\centering
\includegraphics[width=\columnwidth, clip=true]{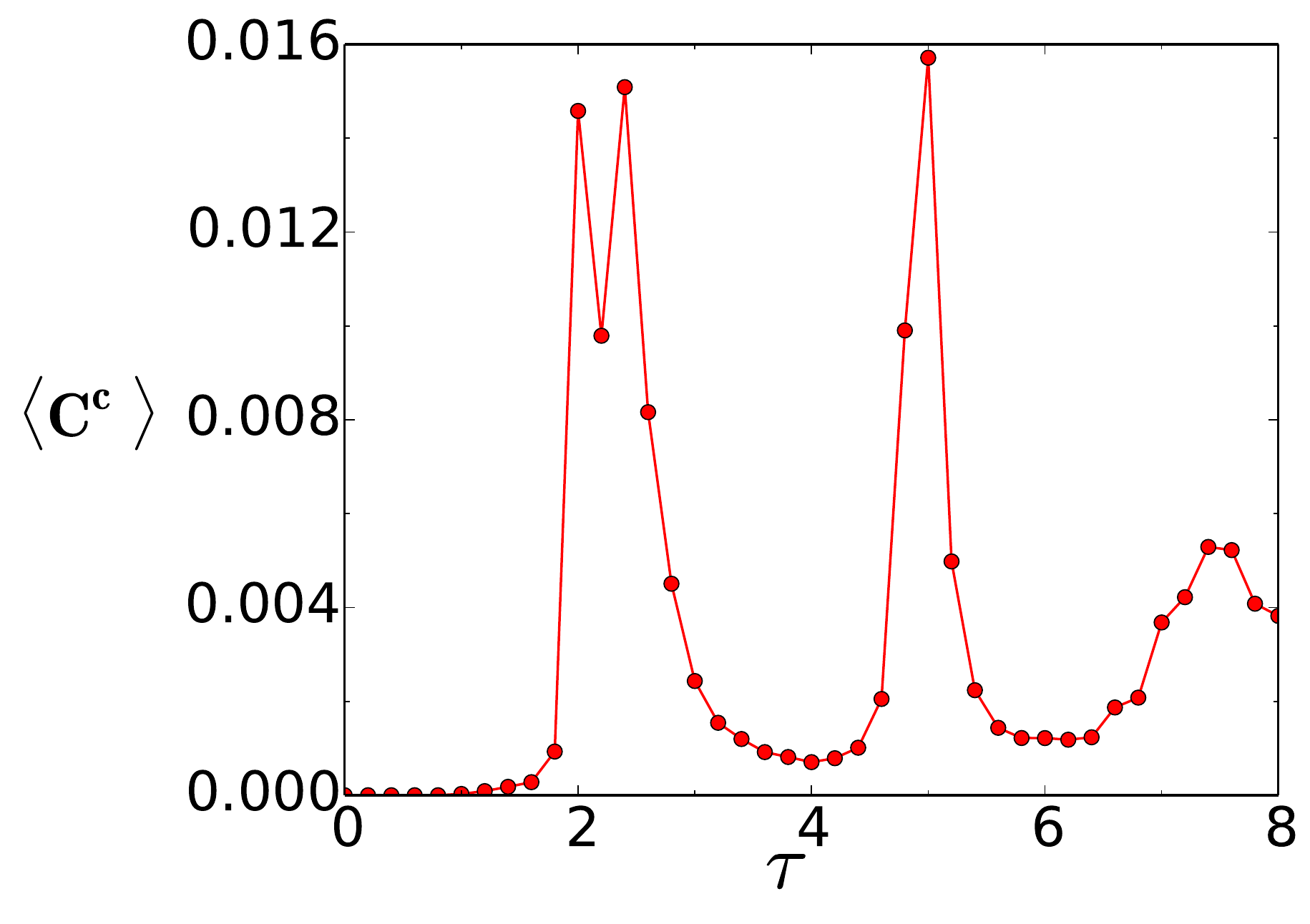}
\caption{Mean clustering coefficient per node $\left< C^c \right>$ as a function of $\tau$ for
the flow networks obtained from the double gyre with $A=0.1$, forced periodically with $2\pi/\omega=5$,
$t_0=0$ and $\epsilon=0.12$.} \label{fig:SpectEps012}
\end{figure}

\begin{figure}
\centering
\includegraphics[width=\columnwidth, clip=true]{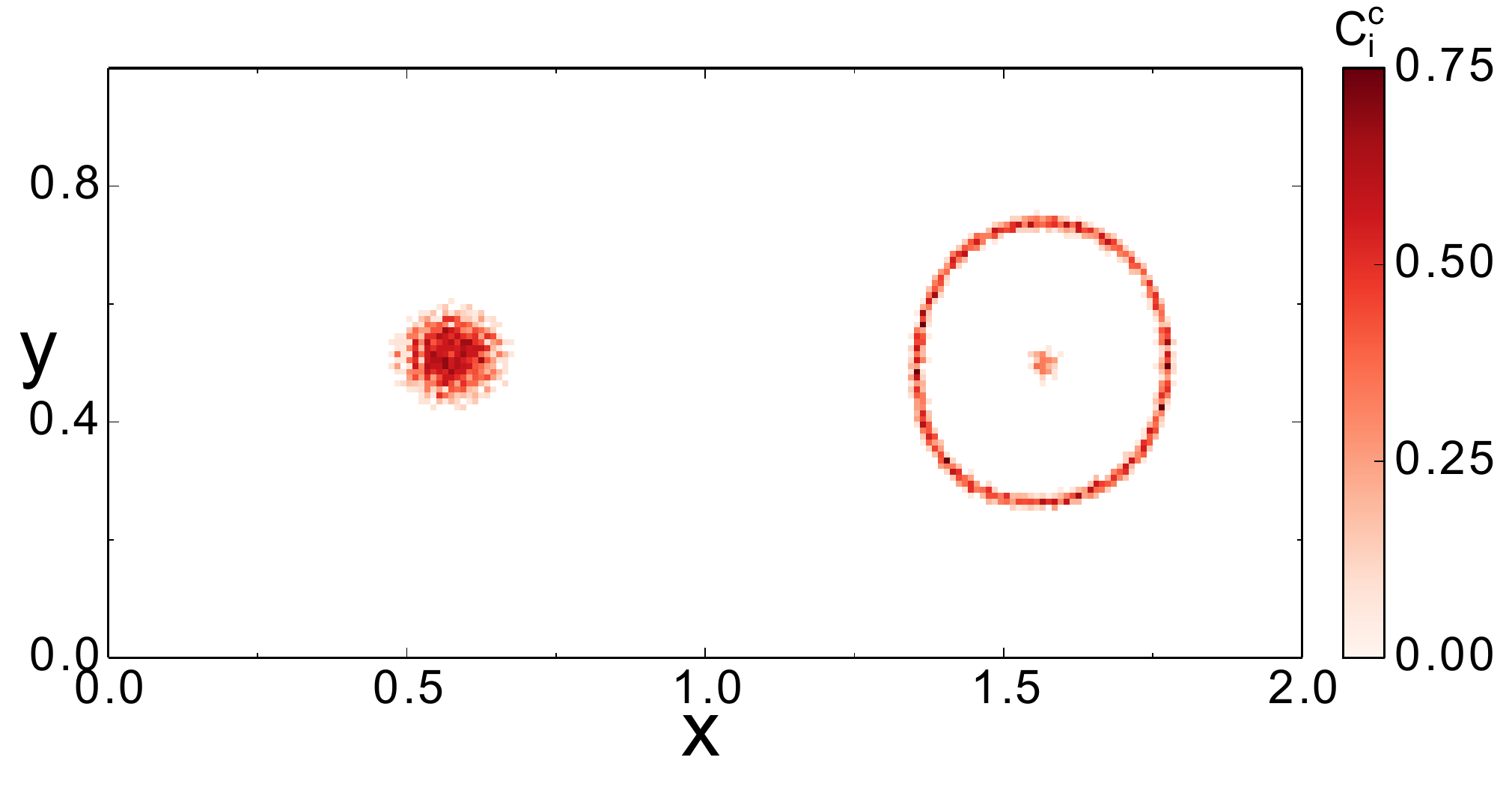}
\caption{Clustering coefficient of the nodes in the flow network of the double gyre with the parameters
of Fig. \ref{fig:SpectEps012} and $\tau=2.2$.}
\label{fig:Eps0p12Tau2p2}
\end{figure}

When the period of the flow does not coincide with the period
$\tau$ used to define the network we are in a quasiperiodic
situation that we expect to be similar to the case of aperiodic
flows: we do not expect exact periodic orbits to be present.
Nevertheless, we can still explore the behavior of the
clustering with changing $\tau$. Again transport of particles
between regions $A$, $B$, $C$ following the pattern $A
\rightarrow B \rightarrow C \rightarrow A$ during the time
interval $[t_0,t_0+\tau]$ will lead to non-vanishing
clustering, although we can not be sure that they will give
periodic time-respecting paths after three iterations since
flow will not be the same in the next intervals
$[t_0+n\tau,t_0+(n+1)\tau]$. They will be approximate periodic
paths if the flow changes slowly on the time scale $\tau$.
Figure \ref{fig:SpectEps012} shows the mean clustering for
fixed $\epsilon=0.12$ and changing $\tau$. There is a sharp
peak at $\tau=5$, as expected, since as shown before there are
cycles $A\rightarrow B \rightarrow C \rightarrow A$ which
indeed lead to periodic trajectories. But we see also high
values of clustering for $2 \lesssim \tau \lesssim 2.6$
indicating that cyclic triangles also occur during the interval
$[t_0,t_0+\tau]$ for these values of $\tau$. Figure
\ref{fig:Eps0p12Tau2p2} shows that, certainly, these regions
containing triangles are well visible in the system for
$\tau=2.2$. Their locations indicate that they are remnants of
the $\epsilon=0$ periodic orbits. They are no longer periodic,
but between $[t_0,t_0+\tau]$ with $t_0=0$ there is cyclic
interchange of particles between these nodes.

\subsection{Steady three-dimensional flow}
\label{subsec:steady3d}

\begin{figure}
\centering
\includegraphics[width=\columnwidth, clip=true]{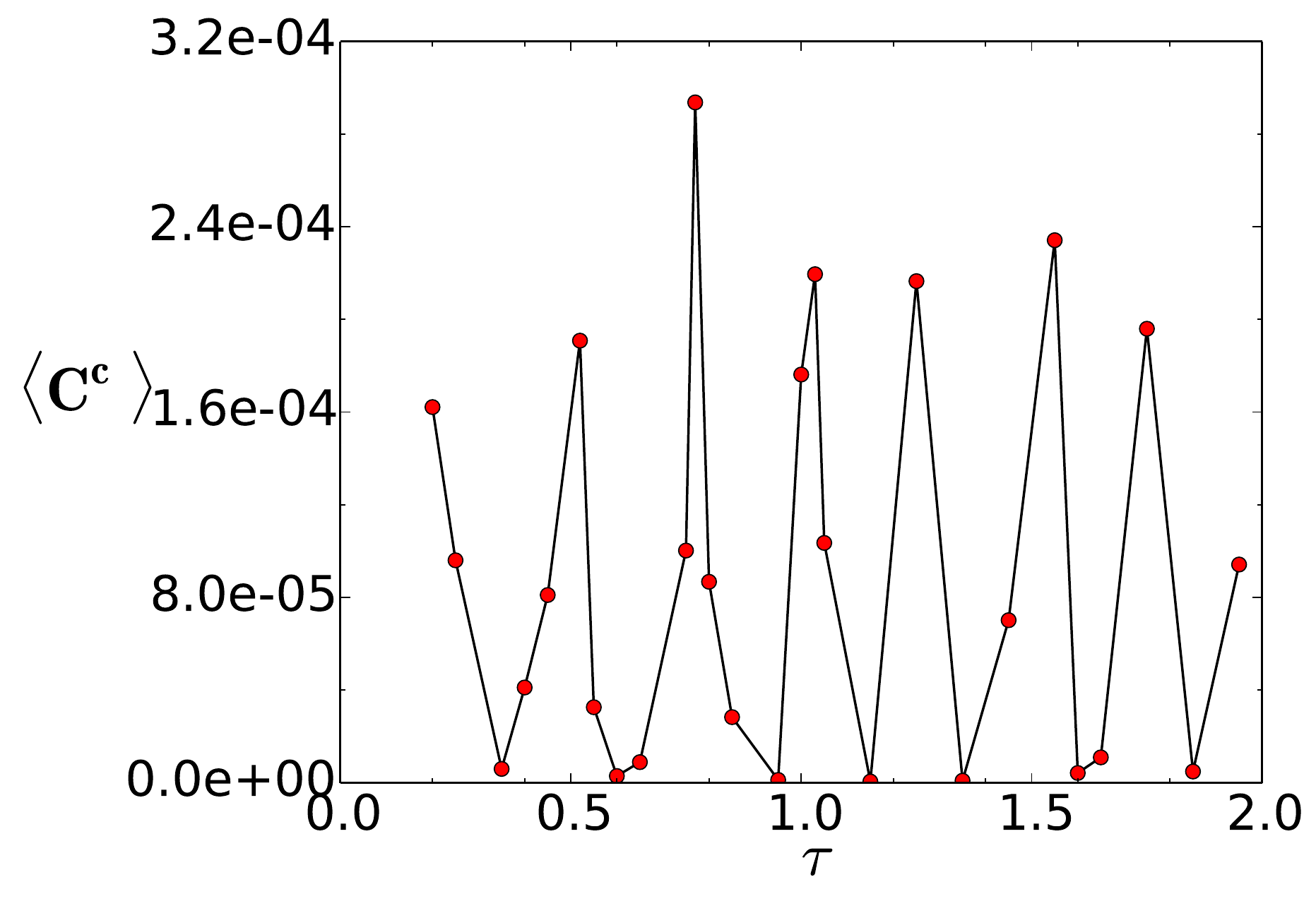}
\caption{Mean clustering coefficient per node $\left< C^c \right>$ as a
function of $\tau$ for the Lorenz system.}
\label{fig:SpectLorenz}
\end{figure}

For steady three-dimensional flows (and in general for steady
$d$-dimensional flows), triangles $A\rightarrow B \rightarrow C
\rightarrow A$ identified during an interval $[t_0,t_0+\tau]$
will remain as triangles in successive intervals thus leading
to periodic orbits of period $3\tau$. This is similar to the
previously studied case of steady 2d flow but, at variance with
it, here the periodic orbits are generically isolated and they
could coexist in phase space with more complex chaotic
trajectories. In order to explore a situation markedly distinct
from the 2d steady double-gyre considered in Sect.
\ref{subsec:steady2d}, which is an incompressible flow, here we
consider a well known dissipative dynamical system, the Lorenz
flow \cite{lorenz1963deterministic,sparrow1982}. The
trajectories $\bx(t)=(x(t),y(t),z(t))$ are the solutions of the
system of equations:
\BA
\dot x &=& s (y-x)\nonumber \\
\dot y &=& r x - y -x z \nonumber \\
\dot z &=& -b z + x y\ .
\label{LorenzEqs}
\EA

We discretize the $(x,y,z) \in
[-20,20]\times[30,30]\times[0,50]$ portion of phase space into
$31616$ cubic boxes, so defining the nodes in our flow network
(each node thus represents a cubic region of lateral size
approximately $1.59$ units). The transport matrix
$\mP(t_0,\tau)$ is computed by releasing 1000 particles from
each of the boxes. We take the well studied parameter set
$s=10$, $b=8/3$, $r=28$ for which the chaotic Lorenz attractor
is well developed and dominates the dynamics. In our initial
condition all boxes have the same density, but under evolution
the particles concentrate in the Lorenz attractor, leaving the
rest of phase space empty. The presence of cyclic triangles is
then expected only in the neighborhood of the attractor.
Embedded in it there is an infinity of (unstable) periodic
orbits of different periods
\cite{sparrow1982,viswanath2003symbolic}. These orbits have
been computed by a variety of techniques
\cite{sparrow1982,viswanath2003symbolic,galiasshort2008} and
are expected to give rise to non-vanishing clustering for the
flow network of $\tau$ given by one third of the corresponding
period. Figure \ref{fig:SpectLorenz} shows the mean clustering
per node for the networks constructed from the Lorenz system at
different values of $\tau$. This provides a kind of
`spectroscopy' by which the presence of orbits of different
periods is revealed. We see peaks at $\tau_1=0.52$,
$\tau_2=0.77$, $\tau_3=1.03$, etc. which correspond very well
(within the step sizes we have used to explore $\tau$) to
one-third of the period of the shortest periodic orbits of the
system for these parameter values
\cite{viswanath2003symbolic,galiasshort2008}: $3\tau_1=1.56$
resonates with the periodic orbit LR (meaning one turn to the
left region of the Lorenz attractor and another turn right,
repeated periodically) of period $T_1\approx 1.5587$. The
resonance at $3\tau_2=2.31$ reveals the LLR and RRL orbits,
both with period $T_2\approx 2.3059$. Next resonance, at
$3\tau_3=3.09$ locates the presence of the orbits LLLR and
RRRL, with $T_3\approx 3.0236$, and LLRR, with $T_4\approx
3.0843$ (our scan in $\tau$ and the node discretization has not
been fine enough to distinguish between them).

\begin{figure*}
\centering
\includegraphics[width=\textwidth, clip=true]{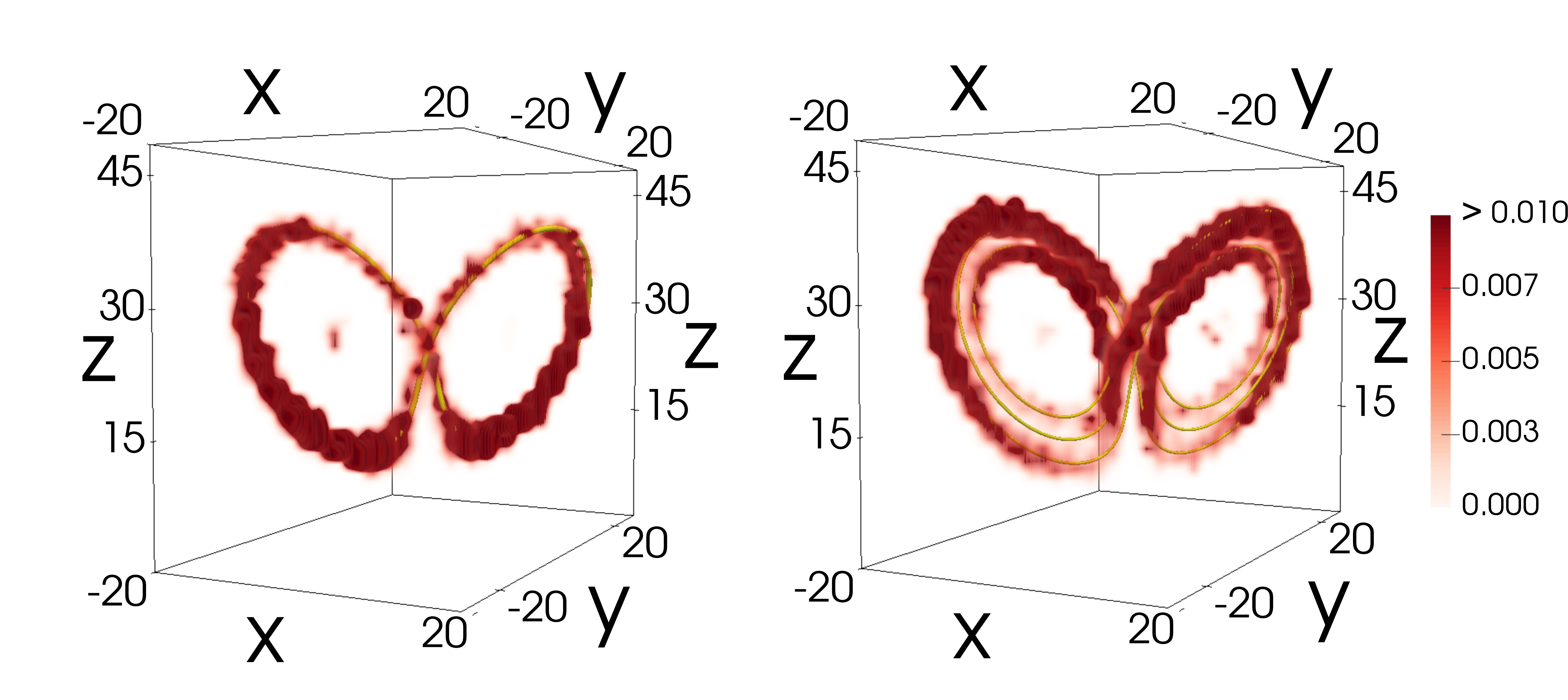}
\caption{Clustering coefficient for the flow network
constructed for the Lorenz model at $\tau=0.52$ (left) and
$\tau=0.77$ (right). Also plotted (yellow thick lines partially
visible inside the red regions) are the periodic orbit LR
(left) and the two periodic orbits LLR and RRL (right). To make
visible all locations with nonvanishing $C_i^c$, nodes with
clustering coefficient larger than the maximum indicated in the
color bar are plotted with the same color.}
\label{fig:LorenzOrbits}
\end{figure*}

To show that certainly the peaks in Fig. \ref{fig:SpectLorenz}
correspond to `resonances' of the value of $\tau$ with these
periodic orbits of period $3\tau$ we show in Fig.
\ref{fig:LorenzOrbits} the clustering coefficient of the
different nodes together with the periodic orbits present at
these values of $\tau$. It is seen how the clustering
coefficient identifies in a coarse-grained way the location of
the periodic orbits.

\subsection{Aperiodic flow}
\label{subsec:aperiodic}

\begin{figure}
\centering
\includegraphics[width=\columnwidth, clip=true]{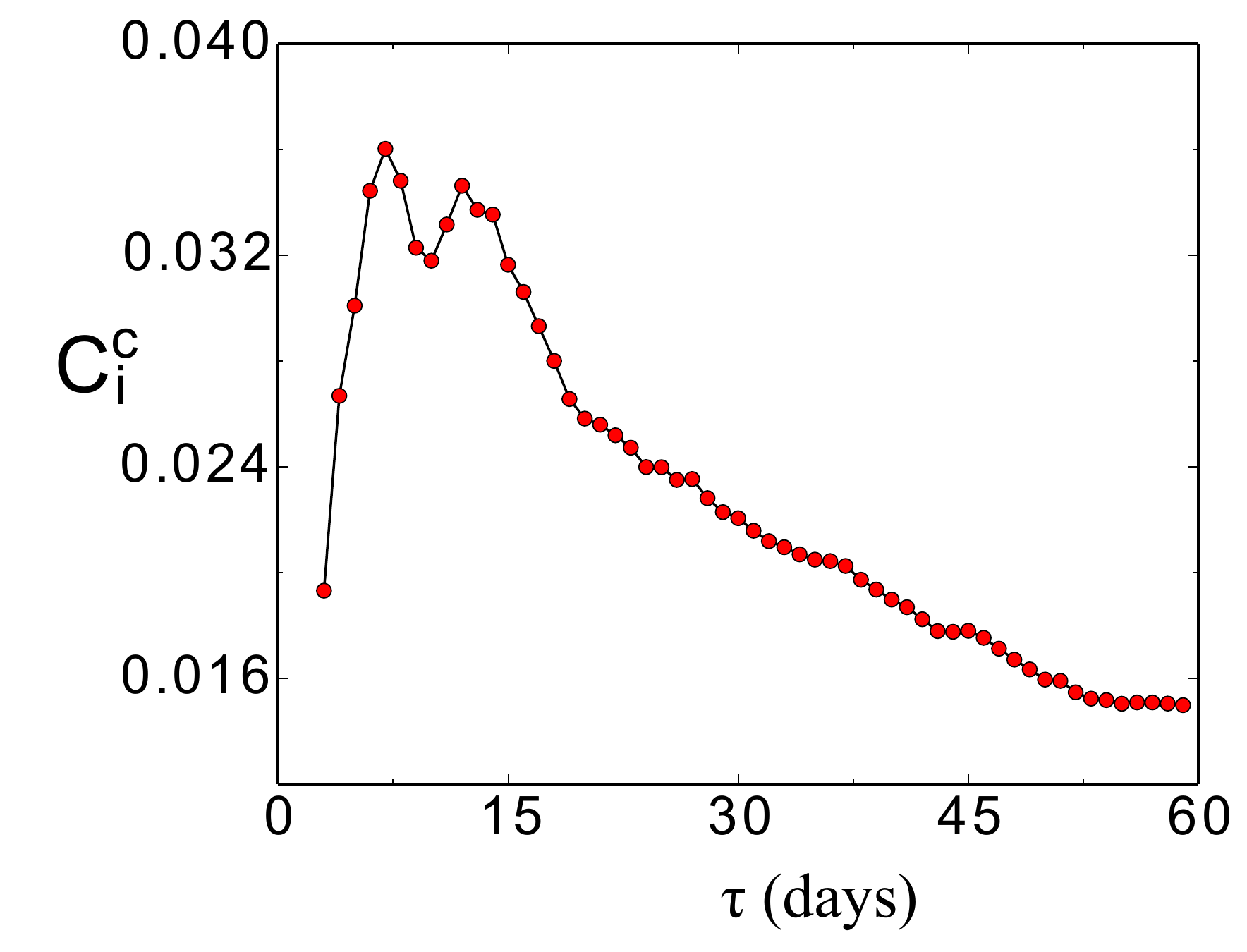}
\caption{Mean clustering coefficient per node $\left< C^c \right>$
as a function of $\tau$ for the flow networks constructed from the MFS surface velocity field
in the Mediterranean starting at $t_0=$ January 1st 2010.}
\label{fig:SpectOcean}
\end{figure}

\begin{figure*}
\centering
\includegraphics[width=\textwidth, clip=true]{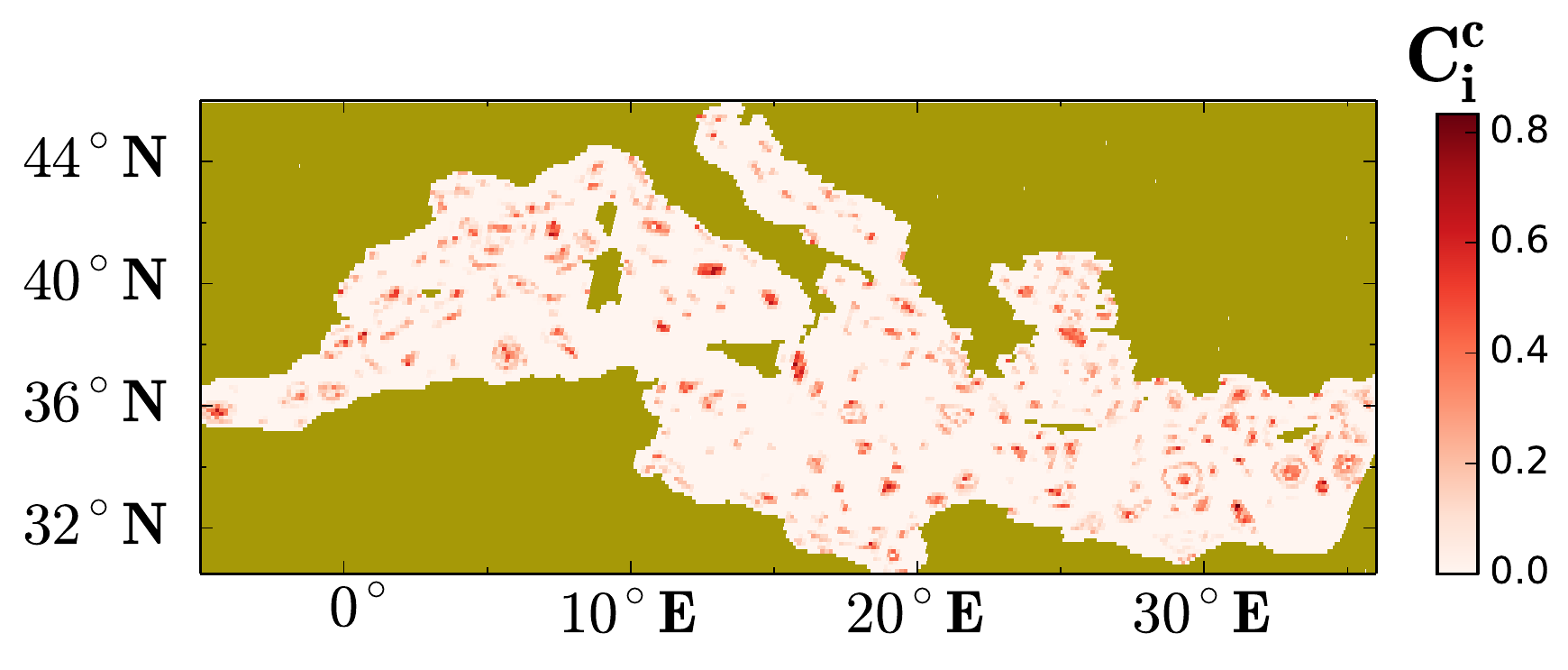}
\caption{Clustering coefficient of the nodes of the flow
network constructed from the MFS surface velocity field in the
Mediterranean with starting date January 1st 2010 and $\tau=6$
days.} \label{fig:Ocean}
\end{figure*}

To explore the behavior of clustering in an aperiodic flow we
use a surface velocity field obtained from the Mediterranean
Forecasting System (MFS), an hydrodynamic model based on
NEMO-OPA (Nucleus for European Modelling of the
Ocean-PArallelis\cite{madec2008nemo}, version 3.2) at
$1/16^\circ$ horizontal resolution and 72 unevenly spaced
vertical levels\cite{oddo2009nested}. We consider the
horizontal motion on the third layer below the surface
corresponding to about 10 meters depth, which gives a good
description of surface transport not excessively affected by
wind stress.

The flow network is constructed
\cite{rossi2014hydrodynamic,sergiacomi2015flow,sergiacomi2015most}
by discretizing the continuous ocean into quasi-square boxes of
$1/8^\circ$ horizontal size using an area-preserving sinusoidal
projection. Each node is filled at $t_0$ with 100 ideal fluid
particles, and their trajectories integrated during a time
$\tau$ to get $\mP(t_0,\tau)$.

Figure \ref{fig:SpectOcean} shows a very broad `resonance' for
$3 \leq \tau \leq 16$ days which identifies potential cyclic
motions of periods $9-48$ days, the range characteristic of
energetic mesoscale structures such as eddies. Figure
\ref{fig:Ocean} displays the spatial distribution of the
clustering coefficient for $\tau=6$ days. The structures and
their locations are reminiscent of ocean eddies, specially the
ones that are semipermanent\cite{millot2005circulation}, such
as the Alboran gyre (just east of Gibraltar strait), the Ionian
anticyclones (east of Sicily) or the Shikmona gyre (south of
Cyprus). Note that several of these objects have a concentric
structure reminiscent of the nested periodic orbits of Fig.
\ref{fig:steady}c,d or \ref{fig:Eps0p12Tau2p2}. For larger
$\tau$ the structures become much weaker (smaller $C^c_i$) but
also larger, suggestive of the big regional gyres in the
Adriatic, the Tyrrhenian , etc. We do not expect perfect
periodic orbits to be present in the turbulent particle motion
produced by this ocean velocity field, but the small and slow
displacements of the semipermanent
gyres\cite{millot2005circulation} detected by the clustering
coefficient (Fig. \ref{fig:Ocean}) makes very likely that they
contain nearly-periodic circulations with approximate period in
a range including $3\tau=18$ days. In fact, these structures
have\cite{millot2005circulation} radii of about 60-150 $km$ and
typical circulations speeds of 0.2-0.6 $m/s$ , leading to
circulation times in the range 7-40 days, consistent with the
periods associated to the broad maximum of Fig.
\ref{fig:SpectOcean}. In any case non-vanishing clustering
reveals cyclic motions in the interval $[t_0,t_0+\tau]$.

\section{Further extensions}
\label{sec:extensions}

\begin{figure}
\centering
\includegraphics[width=\columnwidth, clip=true]{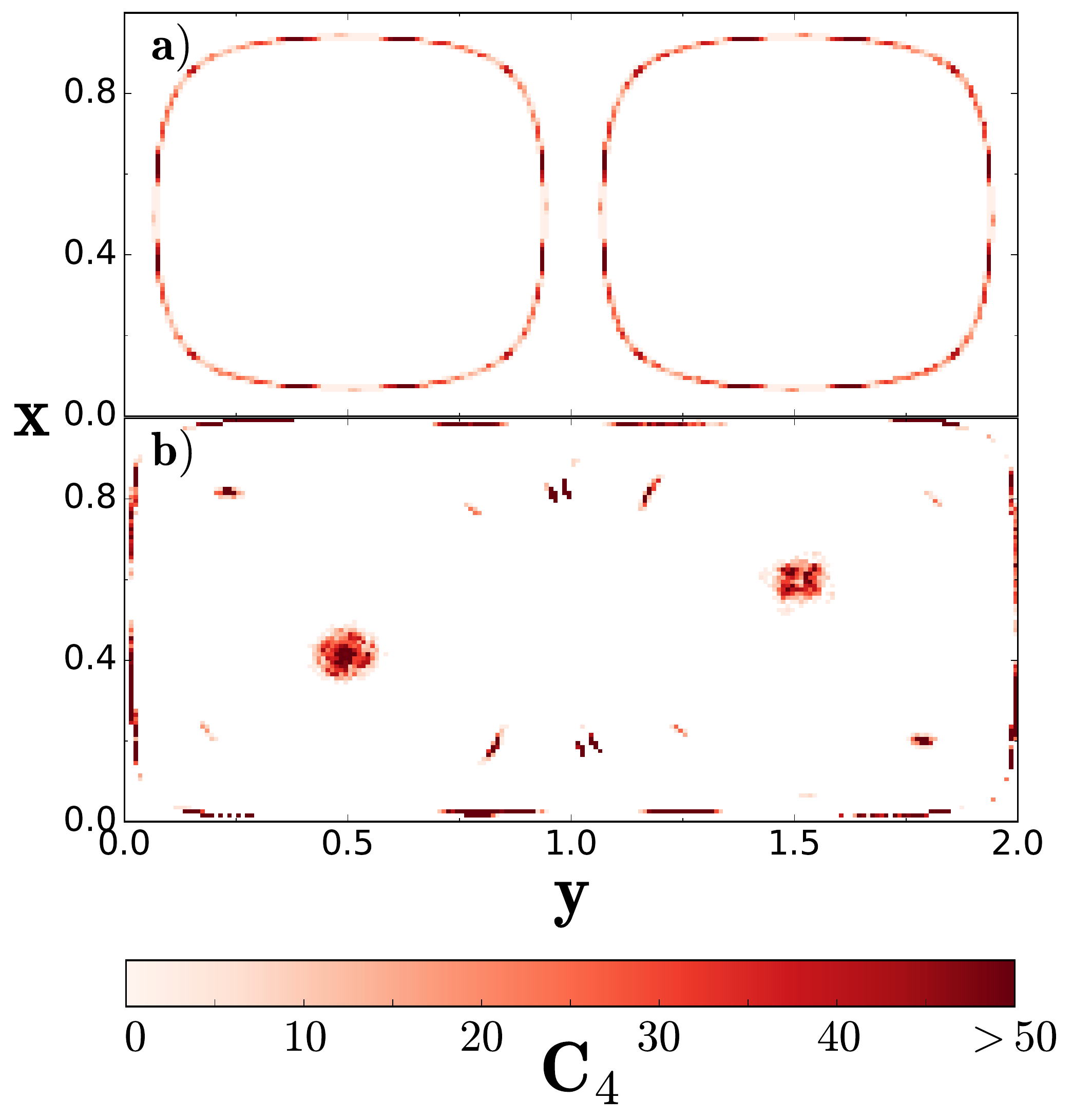}
\caption{Generalized clustering coefficient, $C_4$, counting the number of cyclic polygons
of 4 sides at each node, for the double-gyre flow. a) Steady case, $\epsilon=0$, $\tau=3$;
an orbit with period $4 \tau=12$ is highlighted. b) Periodic case with $\epsilon=0.07$
and forcing period $2\pi/\omega=5=\tau$. }
\label{fig:C4}
\end{figure}

Standard clustering coefficients count triangles. But one can
also count the number of rectangles, pentagons, etc. adjacent
to a given node in a network. We have checked explicitly in the
double-gyre flow example that, as expected, the generalized
cluster coefficients highlight orbits of period $4\tau$,
$5\tau$, etc. An example of this is shown in Fig. \ref{fig:C4}
for the double-gyre flow. We display a generalized clustering
coefficient $C_4$ which is the number of cyclic polygons of 4
sides passing through each node. Panel a), constructed with
$\tau=3$ for the steady case, highlights the same orbit of
period $4\tau=12$ as in Fig. \ref{fig:steady}c. Panel b)
displays $C_4$ for $\epsilon=0.07$ and flow period
$2\pi/\omega$ equal to the integration time $\tau=5$. The
locations of the central period-1 orbits are identified, as
well as four spots around each of them corresponding to the two
elements of elliptic and of hyperbolic period-2 orbits. There
are however other features of difficult interpretation,
apparently associated to the boundaries and to the central
vertical separatrix. Also the central period-1 orbit is not
apparent in Fig. \ref{fig:C4}a). We conclude that the
generalized clustering coefficients contain in principle
similar information to the standard triangular clustering, but
in a more imprecise way arising probably from the extension of
the Markov approximation to longer cycles.

An alternative characterization of triangular flow patterns
which does not take into account the directional nature of flow
can be done with the \emph{undirected} cluster coefficient
\cite{newman2009networks,fagiolo2007clustering}. It is defined
by constructing a new \emph{undirected} network with adjacency
matrix $\mU$ such that $\mU_{ij}=1$ if either $\mA_{ij}=1$ or
$\mA_{ji}=1$, and else $\mU_{ij}=0$. The degree of site $i$ is
now $k_i=\sum_{i\ne j} \mU_{ij}$. The undirected clustering
coefficient $C_i^{u}$
is\cite{newman2009networks,fagiolo2007clustering} the number of
triangles in this new network involving site $i$, over the
total number of possible triangles at that node for such degree
($k_i (k_i-1)/2$):
\BE
C^u_i = \frac{\left(\mU^3\right)_{ii}}{k_i (k_i-1)} \ .
\label{cu}
\EE
We have computed $C^u_i$ for several of the flow examples
considered above. For the double gyre case non-vanishing
$C^u_i$ occurs essentially in the same nodes as non-vanishing
$C^c_i$. This indicates that nearly all triangles present in
double-gyre flow networks are of the cycle type (type a) in
Fig. \ref{fig:triangles}). This may be due to the strong
constraints imposed by the incompressibility of the flow. The
situation for the Lorenz system, however, is completely
different: the higher values of $C^{u}_i$ occur in wide tubes
that intersect the Lorenz attractor transversally, and that
seem to trace portions of the stable manifolds of the
non-vanishing unstable fixed points embedded in the attractor.
We interpret this preliminary observation as revealing an
abundance of triangles of type c) in Fig. \ref{fig:triangles}:
different particles released in node $A$, on a region close to
the stable manifold of one of the fixed points, fall onto the
attractor after a time $\tau$ at slightly different places, $B$
and $C$, which are connected by the dynamics in the attractor.
Further work is needed to establish if this observation is a
general feature of dissipative dynamical systems or
compressible flows, so that different clustering coefficients
of the associated flow network would be useful to locate
complex structures beyond periodic orbits.

We finally mention as another future development the full
consideration of the weighted nature of the flow network, i.e.
using matrix $\mP$ instead of $\mA$. We expect, for example,
that triangle strength would be higher around stable
trajectories as compared to unstable trajectories, since more
fluid particles will remain close to the stable ones.

\section{Conclusions}
\label{sec:conclusions}

In summary, for flow networks with integration time $\tau$
constructed for steady flows or for periodic ones of period
$\tau$ (or submultiple of it), the cyclic clustering
coefficient is non-vanishing in regions locating periodic
orbits of period $3\tau$. Both stable and unstable orbits are
detected. Computation of the transport or adjacency matrices on
which the network approach is based could be expensive (but
easily parallelizable) because of the large number of
Lagrangian particles that should be released. But once
obtained, the matrices can be used to obtain a variety of flow
diagnostics (Lyapunov exponents, community detection, and other
applications mentioned in the Introduction). The computation of
the clustering itself represents just a small computational
overhead that gives access to an important aspect,
periodicities, of flow transport.

Scanning a range of $\tau$ provides a kind of spectroscopic
tool that gives high mean clustering per node at $\tau$ when
periodic orbits of period $3\tau$ exist and have large impact
on the phase space of the system. Even for aperiodic networks
clustering still identifies cyclic triangular paths during the
interval $[0,\tau]$. They could again be interpreted as
periodic trajectories in an approximate way when the time scale
of change of the temporal network is slow compared to $\tau$.
It is likely that a more precise identification of periodic
paths in aperiodic networks could be achieved by replacing the
cyclic clustering coefficient by a \emph{temporal} version of
it, for example by replacing the third power of the single
matrix $\mA$ in Eq. (\ref{cc}) by a product of three adjacency
matrices at consecutive times. This line of work is worth to be
explored in the future.

At variance with standard methods to compute periodic
trajectories in dynamical systems\cite{Nayfeh1995}, clustering
in networks highlights fluid nodes, which are finite regions,
and then it detects thick tubes of trajectories instead of
single pathways. Thus, this network approach can not compete
with standard dynamical systems methods when \emph{precision}
in the location of periodic trajectories is required.
Nevertheless the coarse-graining implied by the network
discretization and the Markov approximation is equivalent to
perturbing trajectories with diffusing
noise\cite{froyland2013analytic} and thus our thick network
paths would be very relevant objects when dealing with
transport and dispersion in the presence of diffusion and/or
noise. More in general, we expect that the connections among
dynamical systems and network theory here proposed would open
additional theoretical developments and fruitful transfer of
concepts and methods across these two fields.

\begin{acknowledgments}
We acknowledge financial support from Spanish Ministerio de
Econom\'{\i}a y Competitividad and Fondo Europeo de Desarrollo
Regional under grants LAOP, CTM2015-66407-P (MINECO/FEDER) and
ESOTECOS projects FIS2015-63628-C2-1-R (MINECO/FEDER), FIS2015-
63628-C2-2-R (MINECO/FEDER). E.S-G. acknowledges partial
support from the program ``Investissements d'Avenir'' launched
by the French Government  and implemented  by ANR under grants
ANR-10-LABX-54 MEMOLIFE  and ANR-11-IDEX-0001-02 PSL* Research
University.
\end{acknowledgments}

\bibliography{LINCreferences,LINC}

\end{document}